\documentclass[12pt]{article}
 \usepackage[paperheight=12in,paperwidth=8.75in]{geometry}

\usepackage{setspace}
\usepackage[T1]{fontenc}
\usepackage{times}
\usepackage{palatino}
\usepackage{lmodern}
\usepackage[latin1]{inputenc}
\usepackage{epsfig}
\usepackage[english]{babel}
\usepackage{color}
\usepackage{graphicx}
\usepackage{dcolumn}
\usepackage{moreverb}
\usepackage{amsmath,amssymb,amsfonts}
\usepackage[all]{xy}
\usepackage{graphicx}

\begin{document}
\numberwithin{equation}{section}
\newcommand{\boxedeqn}[1]{%
  \[\fbox{%
      \addtolength{\linewidth}{-2\fboxsep}%
      \addtolength{\linewidth}{-2\fboxrule}%
      \begin{minipage}{\linewidth}%
      \begin{equation}#1\end{equation}%
      \end{minipage}%
    }\]%
}


\newsavebox{\fmbox}
\newenvironment{fmpage}[1]
     {\begin{lrbox}{\fmbox}\begin{minipage}{#1}}
     {\end{minipage}\end{lrbox}\fbox{\usebox{\fmbox}}}

\raggedbottom
\onecolumn

\newtheorem{th1}{Theorem}[section]
\newtheorem{df}{Definition}[section]
\newtheorem{lem}{Lemma}[section]
\newtheorem{cor}{Corollary}[section]
\newtheorem{pro}{Proposition}[section]
\newtheorem{ex}{Example}[section]
\newtheorem{rem}{Remark}[section]

\parindent 8pt
\parskip 10pt
\baselineskip 16pt
\noindent\title*{{\Large{\textbf{Optical soliton solutions of the Biswas-Arshed model by the $\tan(\circleddash/2)$ expansion approach }}}}
\newline
\newline
Md Fazlul Hoque and Harun-Or-Roshid
\newline
\newline
 Department of Mathematics, Pabna University of Science and Technology, Pabna-6600, Bangladesh
\newline
\newline
E-mail: $fazlul\_math@yahoo.co.in$; $harunorroshidmd@gmail.com $
\newline
\
\begin{abstract}
In this paper, we consider the Biswas-Arshed model (BAM) with nonlinear Kerr and power law. We integrate these nonlinear structures of the BAM to obtain optical exact solitons that passing through the optical fibers. To retrieve the solutions, we apply the $\tan(\circleddash/2)$ expansion integral scheme to the structures of the BAM nonlinearity. The novel solutions present optical shock wave, double periodic optical solitons, interaction between optical periodic wave and optical solitons, and optical periodic and rogue waves for both structures of the model. It is shown that the amplitude of the periodic double solitons waves gradually increases and reached the highest peak at the moment of interaction, and it goes to diminish for a much larger time. In fact, we show that the amplitude of the wave for the interaction between periodic and optical solitons, gradually increases with beat phenomena. To the purpose, all these types of optical solitons can be frequently used to amplify or reduce waves for a certain hight. Moreover, we describe the physical phenomena of the solitons in graphically. 
\end{abstract}


\section{Introduction}
Optical transmission of solitons motion plays a central role in a variety of branches in communication sciences. The general concepts of transmission optical solitons in nonlinear optical fiber systems are fundamentally important in controlling optical continuum and transferring informations for very long distances. The general nonlinear Schrodinger equation can be successfully addressed the picosecond pulses of the wave dynamics involving the coefficient of the group velocity dispersion (GVD) and the self-phase modulation (SPM) \cite{Agr1, Has1}. If ultrashort pulses durations less than 100 femtoseconds, it would be motivated, attractive and desirable to increase the power of high-bit-rate transmission systems, higher-order Kerr dispersion, the slow nonlinear activity, and the third-order dispersion play a vital role in optical fiber media \cite{Agr1, Zha1}.
\\
Due to valuable applications in optical communications and optical signal transmitting systems, the theoretically predicted solitons as well as experimentally study of the dynamical temporal optical solitons were observed in fiber lasers \cite{Son1}. The study of solitons covers as well other areas of physics including plasma physics, fluid dynamics, biological systems and so on. With such effects, there are many classes of nonlinear models such as Lakshmanan-Porsezian-Daniel model, complex Ginzburg-Landau equation, Kundu-Eckhaus equation, Fokas-Ienells equation, Kaup-Newll model, Radhkrishnan-Kundu-Lakshmanan equation, resonant nonlinear Schrodinger's equation, Kadomtsev-Petviashvili-Benjamin -Bona-Mahony, Boiti-Leon-Manna-Pempinelli equation and the references therein \cite{Bis2, Bis3, Mir1, Bis4, Bis5, Bis6, Bis7, Yang1, Liu1, Liu2, Sula1, Tanw1, Mana1, Hoss1}. These models have been successfully addressed dynamical transmission solitons solutions.
\\
In recent years, Biswas and Arshed newly introduced one governing nonlinear model and studied the wave profile of optical solitons in existence of super order dispersions with negligibly SPM \cite{Bis1}. They also pointed out the bright, singular and combo-solitons for the two integration structures of the model. It is fundamentally effective in investigating the dynamical solitons in optical fibers and metameterials in case of both low GVD and nonlinearity. The model known as Biswas-Arshed model (BAM) has been used by several authors to obtain various type optical solitons via the trial solution technique \cite{Yil2}, modified simple equation technique \cite{Yil1}, Kerr and power law nonlinearity \cite{Tah1, Tah2}, mapping method \cite{Reh1}, the extended trial function method \cite{Eki1} and parameter restriction approach \cite{Aou1}. All these approaches have previously been restricted mainly to obtain optical single solitons of the BAM. However, the generalization of the interaction of double optical solitons and rogue type solitons is still an unexplored subject via other existing methods. The interaction of optical nonlinearity solitons often lead to periodic double solitons and rogue waves, and the nature of these solitons is unknown.

The purpose of this paper is to show how we can present optical shock wave, optical double solitons, interaction between optical periodic wave and optical solitons, and optical periodic  and rogue waves for both structures of the model. We also show that how they play to increase or decrease the amplitude of the waves for the solitons in graphically. It is based on the $\tan(\circleddash/2)$ expansion approach \cite{Ilh1, Man1, Man2}. To our knowledge these type of investigations for the BAM is a first step in the study of the $\tan(\circleddash/2)$ expansion method.

\section{Review of the $\tan(\circleddash/2)$ expansion approach}
In the section, we review the $\tan(\circleddash/2)$ expansion approach for the nonlinear evolution equations (NLEEs). These methods are highly effective and algebraic schemes to drive the general optical solitons, periodic solitons and rogue waves solutions, and more deeply understand the properties of the model \cite{Ilh1, Man1}.

\subsection{The $\tan(\circleddash/2)$ expansion approach} \label{Sec2}
This method has been extensively used to a class of NLEEs \cite{Ilh1, Man1} to obtain novel exact solutions. The main algorithm of the approach is as follows:\\
{\bf Step 1} Let us consider the nonlinear evolution equation and the travelling variable transformation in the following form:
\begin{eqnarray}
\Gamma(u_{xx}, u_{xt}, u_{tt}, u_{t},u^n, \dots.)=0,\label{E1}
\end{eqnarray}
where $u(x,t)$ is an unknown function, $\Gamma$ is a polynomial of $u(x,t)$ and its derivatives. The NLEE (\ref{E1}) can be converted to the ordinary differential equation by the transformation relation $\zeta=kx-\omega t$ as follows:
\begin{eqnarray}
\mathcal{H}(\Psi, \Psi', \Psi'', \dots )=0.\label{E2}
\end{eqnarray}
The constant $k$ is the wave number and $\omega$ is the frequency of the wave.
\\
{\bf Step 2} One considers the general form of the trial solution \cite{Ilh1, Man1},
\begin{eqnarray}
\Psi(\circleddash)=\sum_{r=0}^n\mathcal{L}_r \tan^{r}(\frac{\circleddash}{2}),\label{E3}
\end{eqnarray}
where $\mathcal{L}_r$ are free constants to be later calculated, such that $\mathcal{L}_n\neq 0$ and the $\circleddash$ is a function of $\zeta$, which satisfies the condition,
\begin{eqnarray}
\frac{d\circleddash}{d\zeta}=\lambda\sin(\circleddash(\zeta))+\mu \cos(\circleddash(\zeta))  +\nu.\label{E4}
\end{eqnarray}
The condition (\ref{E4}) leads to the following exact solutions:
\begin{itemize}
\item [Set 1:] If $\Lambda = \lambda^2+\mu^2-\nu^2 < 0$, and $\mu-\nu \neq 0$, then  $\circleddash(\zeta)=2\tan^{-1}\left[ \frac{\lambda}{\mu -\nu}-\frac{\sqrt{-\Lambda}}{\mu -\nu}\tan(\frac{\sqrt{-\Lambda}}{2}\bar{\zeta}) \right]$.
\item [Set 2:] If $\Lambda = \lambda^2+\mu^2-\nu^2 > 0$, and $\mu-\nu \neq 0$, then  $\circleddash(\zeta)=2\tan^{-1}\left[ \frac{\lambda}{\mu -\nu}+\frac{\sqrt{\Lambda}}{\mu -\nu}\tanh(\frac{\sqrt{\Lambda}}{2}\bar{\zeta}) \right]$.
\item [Set 3:] If $\Lambda = \lambda^2+\mu^2-\nu^2 > 0$, and $\mu \neq 0$ and $\nu=0$, then \\ $\circleddash(\zeta)=2\tan^{-1}\left[ \frac{\lambda}{\mu}+\frac{\sqrt{\lambda^2+\mu^2}}{\mu}\tanh(\frac{\sqrt{\lambda^2+\mu^2}}{2}\bar{\zeta}) \right]$.
\item [Set 4:] If $\Lambda = \lambda^2+\mu^2-\nu^2 < 0$, and $\mu = 0$ and $\nu \neq 0$, then \\ $\circleddash(\zeta)=2\tan^{-1}\left[ -\frac{\lambda}{\nu}+\frac{\sqrt{\nu^2-\lambda^2}}{\nu}\tan(\frac{\sqrt{\nu^2-\lambda^2}}{2}\bar{\zeta}) \right]$.
\item [Set 5:] If $\Lambda = \lambda^2+\mu^2-\nu^2 > 0$, and $\mu-\nu \neq 0$ and $\lambda = 0$, then \\ $\circleddash(\zeta)=2\tan^{-1}\left[ \sqrt{\frac{\mu+\nu}{\mu-\nu}}\tanh(\frac{\sqrt{\mu^2-\nu^2}}{2}\bar{\zeta}) \right]$.
\item [Set 6:] If $\lambda = 0$, and $\nu = 0$, then  $\circleddash(\zeta)=\tan^{-1}\left[ \frac{e^{2\mu\bar{\zeta}}-1}{e^{2\mu\bar{\zeta}}+1}, \frac{2 e^{\mu\bar{\zeta}}}{e^{2\mu\bar{\zeta}}+1} \right]$.
\item [Set 7:] If $\lambda = 0$, and $\nu = 0$, then  $\circleddash(\zeta)=\tan^{-1}\left[ \frac{e^{2\mu\bar{\zeta}}}{e^{2\mu\bar{\zeta}}+1}, \frac{2 e^{\mu\bar{\zeta}}-1}{e^{2\mu\bar{\zeta}}+1} \right]$.
\item [Set 8:] If $\lambda^2+\mu^2 = \nu^2$, then  $\circleddash(\zeta)=2\tan^{-1}\left[ \frac{\lambda \bar{\zeta}+2}{(\mu-\nu)\bar{\zeta}} \right]$.
\item [Set 9:] If $\lambda = \mu = \nu =k\lambda$, then  $\circleddash(\zeta)=2\tan^{-1}\left[ e^{k\lambda \bar{\zeta}} -1 \right]$.
\item [Set 10:] If $\lambda  = \nu =k\lambda$ and $\mu = -\lambda k$, then  $\circleddash(\zeta)=-2\tan^{-1}\left[\frac{e^{k\lambda \bar{\zeta}}}{e^{k\lambda \bar{\zeta}}-1}  \right]$.
\item [Set 11:] If $\nu  = \lambda$, then  $\circleddash(\zeta)=-2\tan^{-1}\left[\frac{(\lambda+\mu)e^{\mu \bar{\zeta}}-1}{(\lambda-\mu)e^{\mu \bar{\zeta}}-1}  \right]$.
\item [Set 12:] If $\lambda = \nu$, then  $\circleddash(\zeta)=2\tan^{-1}\left[\frac{(\mu+\nu)e^{\mu \bar{\zeta}}+1}{(\mu -\nu)e^{\mu \bar{\zeta}}-1}  \right]$.
\item [Set 13:] If $\nu = -\lambda$, then  $\circleddash(\zeta)=2\tan^{-1}\left[\frac{e^{\mu \bar{\zeta}}+\mu-\lambda}{e^{\mu \bar{\zeta}}-\mu-\lambda} \right]$.
\item [Set 14:] If $\mu = -\nu$, then  $\circleddash(\zeta)=2\tan^{-1}\left[\frac{\lambda e^{\lambda \bar{\zeta}}}{1- \nu e^{\lambda \bar{\zeta}}} \right]$.
\item [Set 15:] If $\mu = 0$ and $\lambda = \nu$, then  $\circleddash(\zeta)=- 2\tan^{-1}\left[\frac{\nu \bar{\zeta}+2}{\nu \bar{\zeta}} \right]$.
\item [Set 16:] If $\lambda = 0$ and $\mu = \nu$, then  $\circleddash(\zeta)= 2\tan^{-1}\left[\nu \bar{\zeta} \right]$.
\item [Set 17:] If $\lambda = 0$ and $\mu = -\nu$, then  $\circleddash(\zeta)= -2\tan^{-1}\left[\frac{1}{\nu \bar{\zeta}} \right]$.
\item [Set 18:] If $\lambda = 0$ and $\mu = 0$, then  $\circleddash(\zeta)= \nu \zeta +\mathcal{K} $.
\item [Set 19:] If $\mu = \nu$, then  $\circleddash(\zeta)= 2\tan^{-1}\left[\frac{e^{\lambda \bar{\zeta}}-\nu}{\lambda}  \right]$, 
\end{itemize}
where $\bar{\zeta}=\zeta+\mathcal{K}$, where $\mathcal{L}_0, \mathcal{L}_r$ are constants to be later calculated.
\\
{\bf Step 3.} Inserting the above solutions into (\ref{E3}) together with (\ref{E4}) and substituting into (\ref{E2}), and then solving the algebraic equations for the coefficients $\mathcal{L}_0, \mathcal{L}_r$ and $\omega$. Finally, putting the results into (\ref{E3}) with the above set of solutions, they provide the new exact solutions of the NLEEs (\ref{E1}). In the following sections, we apply this approach to the first and second structures of the Biswas-Arshed model.

\section{First structure of the Biswas-Arshed model}
In this section, we consider the first structure of the Biswas-Arshed model (BAM) \cite{Bis1} to obtain the optical double solitons and the rogue waves via the $\tan(\circleddash/2)$ expansion approach. The first structure of the BAM is \cite{Bis1, Tah1},
 \begin{eqnarray}
i\mathcal{V}_t+ a_1\mathcal{V}_{xx}+a_2\mathcal{V}_{xt}+i(b_1\mathcal{V}_{xxx}+b_2\mathcal{V}_{xxt})=i[\varepsilon(|\mathcal{V}|^2\mathcal{V})_x+\sigma(|\mathcal{V}|^2)_x\mathcal{V}+ \vartheta |\mathcal{V}|^2\mathcal{V}_x].\label{MOD1}
\end{eqnarray}  
 The function $\mathcal{V}(x,t)$ in the model represents the wave profile of soliton, $a_1$ is the coefficient of the GVD, $a_2$ is the spatio-temporal dispersion, $b_1$ is the coefficient of third-order dispersion, $b_2$ is the coefficient of spatio-temporal third-order dispersion, and $\varepsilon$ is the effect of self-steepening, and $\sigma, \vartheta$ are the effect of dispersions. This model was first introduced by Biswas-Arshed \cite{Bis1}, known as the so-called BAM, and it was shown the optical solitons in the presence of higher-order dispersions with negligibly self-phase modulation. Such a model with optical transmission wave solutions has also found in the context of Kerr and non-Kerr law media \cite{Tah1}. This model considered here could have wider applicability and other aspects as optical double solitons, periodic solitons and rogue type wave solutions could be investigated via the $\tan(\circleddash/2)$ expansion method. This method makes the Biswas-Arshed model to be highly interesting. In this section, we apply the transformation variable to (\ref{E1})  
 
 \subsection{The ODE of the first structure of BAM}
 Let us consider the transformation variable \cite{Bis1},
 \begin{eqnarray}
 \mathcal{V}(x,t)=\mathcal{U}(\zeta)e^{i\phi(x,t)},\label{TRV1}
 \end{eqnarray}
 where $\mathcal{U}(\zeta)$ is the amplitude portion with $\zeta=x-\delta t$, the phase component $\phi(x,t)=-k x+w t+\rho$, and the constants $\delta, \rho, k, w$ are respectively, the soliton velocity, the phase constant, the wave number and the frequency of the soliton. After a long computation and an integration, one can present the differential equation \cite{Bis1, Tah1} of (\ref{MOD1}),
 \begin{eqnarray}
 \mathcal{A} \mathcal{U}''-\mathcal{B}\mathcal{U}-\mathcal{C}\mathcal{U}^3=0,\label{OD1}
 \end{eqnarray}
 where $\mathcal{A}=a_1-a_2\delta+3b_1 k-2b_2 \delta k-b_2 w$, $\mathcal{B}=a_1 k^2-a_2 wk+b_1 k^3-b_2 w k^2+w$ and $\mathcal{C}=k(\varepsilon +\vartheta)$. In the following section, we apply the extension $\tan(\circleddash/2)$ method to (\ref{OD1}) and obtain the exact soliton solutions of (\ref{MOD1}).

 \subsection{Application of the $\tan(\circleddash/2)$ method to the first structure of BAM}
\begin{figure}[b]
\begin{center}
\includegraphics[width=5cm, height=5cm]{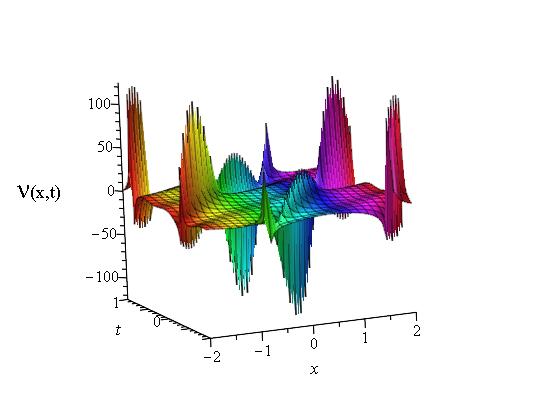}\qquad\qquad \includegraphics[width=6cm, height=4cm]{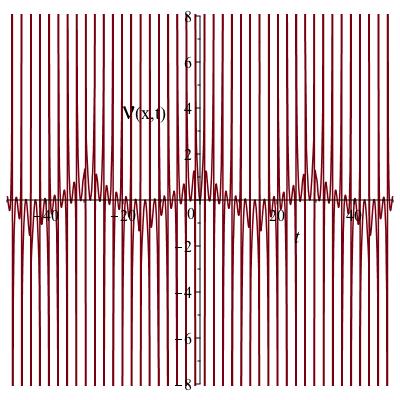}
\caption{The 2D and 3D double periodic  solitons of $\mathcal{V}_{1,1}(x,t)$, for the real  part of $\mathcal{V}_{1,1}(x,t)$ with $a_1=b_1=b_2=\delta=\lambda=\mu=k=1$, $a_2=2$, and $\nu=3$, $\varepsilon=0$.} \label{fig:figure1}
\end{center}
\end{figure}

\begin{figure}[b]
\begin{center}
\includegraphics[width=5cm, height=4cm]{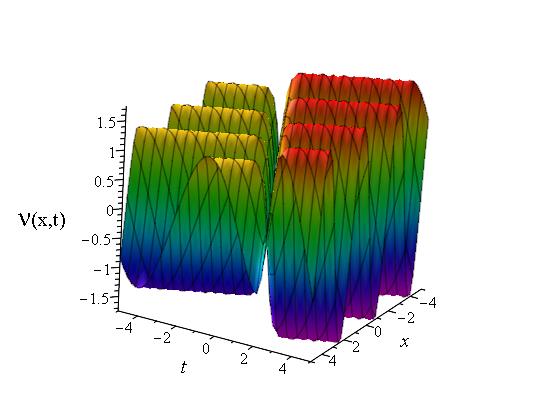}\qquad\qquad \includegraphics[width=4cm, height=4cm]{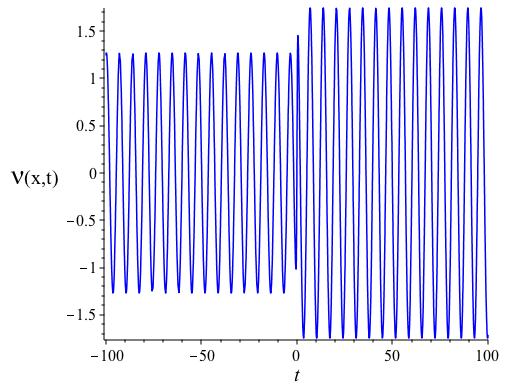}
\caption{The 3D periodic double solitons of $\mathcal{V}_{1,3}(x,t)$, for the imaginary part of $\mathcal{V}_{1,3}(x,t)$ with $a_1=b_1=b_2=\delta=\lambda=k=1$, $a_2=2$, $\mu=3$ and $\nu=1$, $\varepsilon=0$.} \label{fig:figure2}
\end{center}
\end{figure}

\begin{figure}[b]
\begin{center}
\includegraphics[width=5cm, height=5cm]{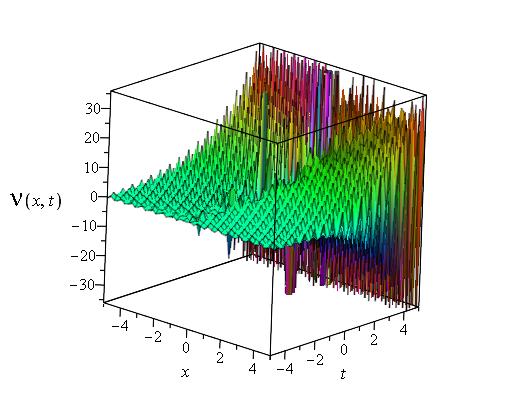}\qquad\qquad \includegraphics[width=5cm, height=5cm]{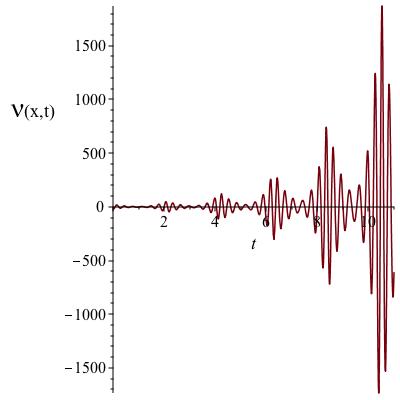} \includegraphics[width=6cm, height=4cm]{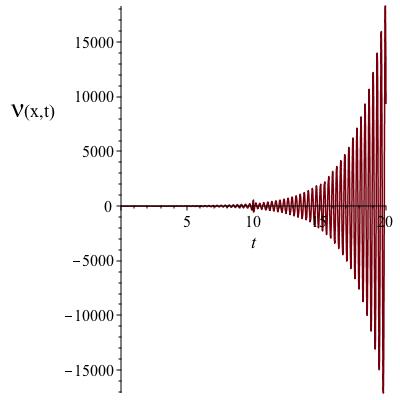}
\caption{The 2D and 3D periodic double solitons of $\mathcal{V}_{1,6}(x,t)$, for the real part of $\mathcal{V}_{1,6}(x,t)$ with $a_1=0.5$ $a_2=b_1=b_2=\lambda=\nu=1$, $\delta=\sqrt{-1}$, $\mu=3$, $k=20$, $\varepsilon=0$ and $\vartheta=0.3$.} \label{fig:figure3}
\end{center}
\end{figure}

\begin{figure}[b]
\begin{center}
\includegraphics[width=5cm, height=5cm]{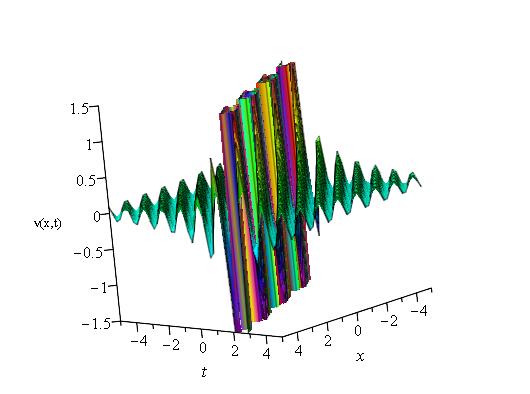}\qquad\qquad \includegraphics[width=5cm, height=5cm]{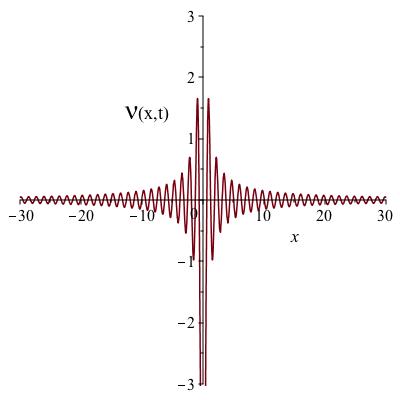}
\caption{The 2D and 3D periodic double solitons of $\mathcal{V}_{1,8}(x,t)$, for the real part of $\mathcal{V}_{1,8}(x,y)$ with $a_1=a_2=b_1=b_2=\lambda=\mu=1$, $a_2=\delta=2$, $k=5$, $\vartheta=2$, $\varepsilon=0$ and $\nu=2$.} \label{fig:figure4}
\end{center}
\end{figure}

 We now compute the balance number of (\ref{OD1}) from power balance of the highest derivative with the highest-order nonlinear terms \cite{Bis1}, that leads to $n=1$. Then the trial solution (\ref{E3}) in the $\tan(\circleddash/2)$ expansion method takes the form,
 \begin{eqnarray}
\Psi(\circleddash)=\mathcal{L}_0 +\mathcal{L}_1 \tan(\frac{\circleddash}{2}).\label{E5}
\end{eqnarray}
Putting (\ref{E5}) into (\ref{OD1}) along with (\ref{E4}), we obtain a polynomial of $\sin(\zeta)$ and $\cos(\zeta)$ functions, whose equating coefficients lead to a system of equations, and the solutions of the system of equations via symbolic computation, yield the following constraints:
{\scriptsize{
\begin{eqnarray}
 &&  w=\frac{(\lambda^2+\mu^2-\nu^2)(a_1-\delta a_2-2\delta k b_2 +3kb_1)+2b_1 k^3+2a_1 k^2}{2 k^2 b_2+\lambda^2 b_2+\mu^2 b_2-\nu^2 b_2+2 k a_2-2}, \nonumber\\&&
 \mathcal{L}_0 = \pm \lambda\sqrt{\frac{2k^3b_1b_2+3k^2a_2b_1+ka_1a_2-3kb_1-a_1-\delta(2k^3b_2^2+3k_1^2a_2b_2+ka_2^2-2kb_2-a_2)}{k(\varepsilon+\vartheta) (2k^2b_2+\lambda^2b_2+\mu^2b_2-\nu^2b_2+2ka_2-2)}}, \nonumber\\&&
 \mathcal{L}_1= \pm (\nu-\mu)\sqrt{\frac{\delta(2k^3b_2^2+3k_1^2a_2b_2+ka_2^2-2kb_2-a_2) -2k^3b_1b_2-3k^2a_2b_1-ka_1a_2+3kb_1+a_1}{k(\varepsilon+\vartheta)(2k^2b_2+\lambda^2b_2+\mu^2b_2-\nu^2b_2+2ka_2-2)}}.\label{Cont1}
 \end{eqnarray}
 }}
Now if we combine the above constraints  with (\ref{E5}), and substituting into (\ref{TRV1}), and the solutions Set [1-19] of (\ref{E4}), we obtain the following seventeen valid exact soliton solutions of (\ref{MOD1}), 
\\
 $\mathcal{V}_{1,1}(x,t)=\left\{\mathcal{L}_0+\mathcal{L}_1\left[ \frac{\lambda}{\mu -\nu}-\frac{\sqrt{-\Lambda}}{\mu -\nu}\tan(\frac{\sqrt{-\Lambda}}{2}\bar{\zeta}) \right]\right\}e^{i(-kx+wt+\varepsilon)} $,
\\
$\mathcal{V}_{1,2}(x,t)=\left\{\mathcal{L}_0+\mathcal{L}_1\left[ \frac{\lambda}{\mu -\nu}+\frac{\sqrt{\Lambda}}{\mu -\nu}\tanh(\frac{\sqrt{\Lambda}}{2}\bar{\zeta}) \right]\right\}e^{i(-kx+wt+\varepsilon)} $,
\\
$\mathcal{V}_{1,3}(x,t)=\left\{\mathcal{L}_0+\mathcal{L}_1\left[ \frac{\lambda}{\mu}+\frac{\sqrt{\lambda^2+\mu^2}}{\mu}\tanh(\frac{\sqrt{\lambda^2+\mu^2}}{2}\bar{\zeta}) \right]\right\}e^{i(-kx+wt+\varepsilon)} $,
\\
$\mathcal{V}_{1,4}(x,t)=\left\{\mathcal{L}_0+\mathcal{L}_1\left[ -\frac{\lambda}{\nu}+\frac{\sqrt{\nu^2-\lambda^2}}{\nu}\tan(\frac{\sqrt{\nu^2-\lambda^2}}{2}\bar{\zeta}) \right]\right\}e^{i(-kx+wt+\varepsilon)}$,
\\
$\mathcal{V}_{1,5}(x,t)=\left\{\mathcal{L}_0+\mathcal{L}_1 \left[ \sqrt{\frac{\mu+\nu}{\mu-\nu}}\tanh(\frac{\sqrt{\mu^2-\nu^2}}{2}\bar{\zeta}) \right]\right\}e^{i(-kx+wt+\varepsilon)}$,
\\
$\mathcal{V}_{1,6}(x,t)=\left\{\mathcal{L}_0+\mathcal{L}_1\tan \left( \frac{1}{2}\tan^{-1}\left[ \frac{e^{2\mu\bar{\zeta}}-1}{e^{2\mu\bar{\zeta}}+1}, \frac{2 e^{\mu\bar{\zeta}}}{e^{2\mu\bar{\zeta}}+1} \right]\right)\right\}e^{i(-kx+wt+\varepsilon)}$,
\\
$\mathcal{V}_{1,7}(x,t)=\left\{\mathcal{L}_0+\mathcal{L}_1\tan \left( \frac{1}{2}\tan^{-1}\left[ \frac{e^{2\mu\bar{\zeta}}}{e^{2\mu\bar{\zeta}}+1}, \frac{2 e^{\mu\bar{\zeta}}-1}{e^{2\mu\bar{\zeta}}+1} \right]\right)\right\}e^{i(-kx+wt+\varepsilon)}$,
\\
$\mathcal{V}_{1,8}(x,t)=\left\{\mathcal{L}_0+\mathcal{L}_1\left[ \frac{\lambda \bar{\zeta}+2}{(\mu-\nu)\bar{\zeta}} \right]\right\}e^{i(-kx+wt+\varepsilon)}$,
\\
$\mathcal{V}_{1,11}(x,t)=\left\{\mathcal{L}_0+\mathcal{L}_1\tan \left( -\tan^{-1}\left[\frac{(\lambda+\mu)e^{\mu \bar{\zeta}}-1}{(\lambda-\mu)e^{\mu \bar{\zeta}}-1}  \right]\right)\right\}e^{i(-kx+wt+\varepsilon)}$,
\\
$\mathcal{V}_{1,12}(x,t)=\left\{\mathcal{L}_0+\mathcal{L}_1\left[ \frac{(\mu+\nu)e^{\mu \bar{\zeta}}+1}{(\mu -\nu)e^{\mu \bar{\zeta}}-1} \right]\right\}e^{i(-kx+wt+\varepsilon)}$,
\\
$\mathcal{V}_{1,13}(x,t)=\left\{\mathcal{L}_0+\mathcal{L}_1\left[ \frac{e^{\mu \bar{\zeta}}+\mu-\lambda}{e^{\mu \bar{\zeta}}-\mu-\lambda}\right]\right\}e^{i(-kx+wt+\varepsilon)}$,
\\
$\mathcal{V}_{1,14}(x,t)=\left\{\mathcal{L}_0+\mathcal{L}_1\left[ \frac{\lambda e^{\lambda \bar{\zeta}}}{1- \nu e^{\lambda \bar{\zeta}}}\right]\right\}e^{i(-kx+wt+\varepsilon)}$,
\\
$\mathcal{V}_{1,15}(x,t)=\left\{\mathcal{L}_0+\mathcal{L}_1\tan \left( -\tan^{-1}\left[\frac{\nu \bar{\zeta}+2}{\nu \bar{\zeta}} \right]\right)\right\}e^{i(-kx+wt+\varepsilon)}$,
\\
$\mathcal{V}_{1,16}(x,t)=\left\{\mathcal{L}_0+\mathcal{L}_1\left[ \nu \bar{\zeta} \right]\right\}e^{i(-kx+wt+\varepsilon)}$,
\\
$\mathcal{V}_{1,17}(x,t)=\left\{\mathcal{L}_0+\mathcal{L}_1\tan \left( -\tan^{-1}\left[\frac{1}{\nu \bar{\zeta}} \right]\right)\right\}e^{i(-kx+wt+\varepsilon)}$,
\\
$\mathcal{V}_{1,18}(x,t)=\left\{\mathcal{L}_0+\mathcal{L}_1\tan \left[\frac{\nu \zeta +\mathcal{K}}{2} \right]\right\}e^{i(-kx+wt+\varepsilon)}$,
\\
$\mathcal{V}_{1,19}(x,t)=\left\{\mathcal{L}_0+\mathcal{L}_1\left[ \frac{e^{\lambda \bar{\zeta}}-\nu}{\lambda} \right]\right\}e^{i(-kx+wt+\varepsilon)}$,\\
where $w$, $\mathcal{L}_0$ and $\mathcal{L}_1$ come from (\ref{Cont1}). There is a note that the solutions $\mathcal{V}_{1,9}(x,t)$ and $\mathcal{V}_{1,10}(x,t)$ are invalid for the constraints.
\newline
\newline
To the graphical description, it is shown that $\mathcal{V}_{1,1}(x,t)$, $\mathcal{V}_{1,4}(x,t)$, $\mathcal{V}_{1,18}(x,t)$ solutions provide double periodic optical solitons. In particular, we present the $2D$ and 3D graphs in Figure \ref{fig:figure1} for the solution $\mathcal{V}_{1,1}(x,t)$; and the solutions  $\mathcal{V}_{1,4}(x,t)$ and $\mathcal{V}_{1,18}(x,t)$ have similar dynamical characteristics to the Figure \ref{fig:figure1}.
\\
The solutions $\mathcal{V}_{1,2}(x,t)$, $\mathcal{V}_{1,3}(x,t)$ and  $\mathcal{V}_{1,5}(x,t)$ represent interaction between periodic waves and solitons, which produce an optical double solitons. We present graphs in Figure \ref{fig:figure2} for a particular solution $\mathcal{V}_{1,3}(x,t)$. It is shown that the amplitude of the solitons increases after the interaction ($t=0$). These types of optical solitons can be extensively used to amplify waves for a certain hight.
\\
The solutions $\mathcal{V}_{1,6}(x,t)$, $\mathcal{V}_{1,7}(x,t)$,  $\mathcal{V}_{1,11}(x,t)$, $\mathcal{V}_{1,12}(x,t)$, $\mathcal{V}_{1,13}(x,t)$, $\mathcal{V}_{1,14}(x,t)$ and $\mathcal{V}_{1,19}(x,t)$ provide to similar interaction between periodic waves and optical solitons. We give graphs in Figure \ref{fig:figure3} for the solution $\mathcal{V}_{1,6}(x,t)$. The graphs show that the amplitude of the optical solitons gradually increases with beat phenomena. 
\\
The solutions $\mathcal{V}_{1,8}(x,t)$,  $\mathcal{V}_{1,15}(x,t)$, $\mathcal{V}_{1,16}(x,t)$,  and $\mathcal{V}_{1,17}(x,t)$ provide to similar interaction between periodic waves and optical solitons. We present a periodic  double solitons for the solution $\mathcal{V}_{1,8}(x,t)$ in Figure \ref{fig:figure4}. The graphs show that the amplitude of the wave gradually increases and reaches highest peak at the moment of interaction, and then the amplitude goes to diminish for a larger time. As it is presented various interactions of optical solitons, and such interactions are stable localized wave packets that can travel large distance in optical fibers remaining their structures \cite{Son1}. Thus these types of interactions solitons would be interested in optical fiber communications.

\section{The second structure of Biswas-Arshed model}
Let us consider the second structure of Biswas-Arshed model \cite{Bis1},
 \begin{eqnarray}
i\mathcal{V}_t+ a_1\mathcal{V}_{xx}+a_2\mathcal{V}_{xt}+i(b_1\mathcal{V}_{xxx}+b_2\mathcal{V}_{xxt})=i[\varepsilon(|\mathcal{V}|^{2q}\mathcal{V})_x+\sigma(|\mathcal{V}|^{2q})_x\mathcal{V}+ \vartheta |\mathcal{V}|^{2q}\mathcal{V}_x],\label{MODS2}
\end{eqnarray} 
where $q$ leads to the nonlinearity of the model. In this section, we apply the $\tan(\circleddash/2)$ expansion approach to this model to investigate the more general exact soliton solutions in different aspects. Making the identification $q=1$, the model (\ref{MODS2}) becomes the first structure of BAM (\ref{MOD1}).
 
\subsection{The ODE of the second structure of BAM}
Using the transformation (\ref{TRV1}) into the model (\ref{MODS2}) leads to the following ODE \cite{Bis1},
 \begin{eqnarray}
 \mathcal{A} \mathcal{U}''-\mathcal{B}\mathcal{U}-\mathcal{C}\mathcal{U}^{2q+1}=0,\label{ODS2}
 \end{eqnarray}
 where $\mathcal{A}=a_1-a_2\delta+3b_1 k-2b_2 \delta k-b_2 w$, $\mathcal{B}=a_1 k^2-a_2 wk+b_1 k^3-b_2 w k^2+w$ and $\mathcal{C}=k(\varepsilon +\vartheta)$. After scaling the ODE (\ref{ODS2}) by $\mathcal{U}=\mathcal{T}^{\frac{1}{2q}}$ yields,
 \begin{eqnarray}
 \mathcal{A}\{(1-2q)(\mathcal{T}')^2+2q \mathcal{T}\mathcal{T}''\}-4q^2\mathcal{B}\mathcal{T}^2-4q^2\mathcal{C}\mathcal{T}^3=0. \label{ODESS2}
\end{eqnarray}

\subsection{Application of the expansion of $\tan(\circleddash/2)$ approach to the second structure of BAM}
\begin{figure}[b]
\begin{center}
\includegraphics[width=6cm, height=6cm]{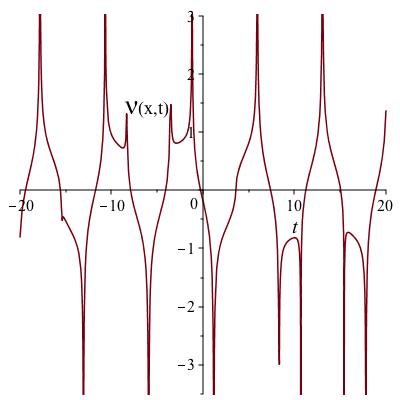}\qquad\qquad \includegraphics[width=6cm, height=6cm]{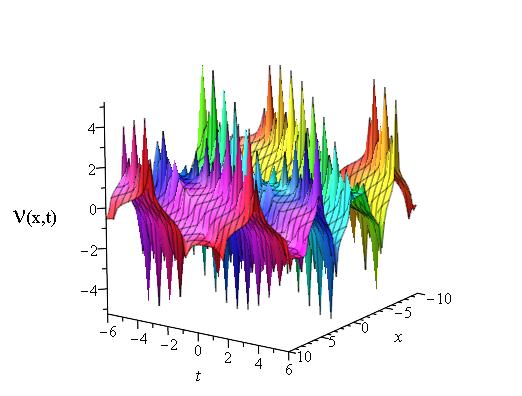}
\caption{The 2D and 3D double periodic solitons of $\mathcal{V}_{2,1}(x,t)$, for the real part of $\mathcal{V}_{2,1}(x,t)$ with $a_1=b_1=b_2=\delta=k=\vartheta=\lambda=1$, $a_2=q=2$, $\mu=1$, $\varepsilon=0$ and $\nu=3$.} \label{fig:figure5}
\end{center}
\end{figure}

\begin{figure}[b]
\begin{center}
\includegraphics[width=6cm, height=6cm]{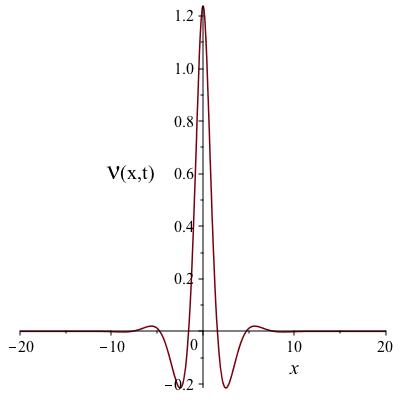}\qquad\qquad \includegraphics[width=6cm, height=6cm]{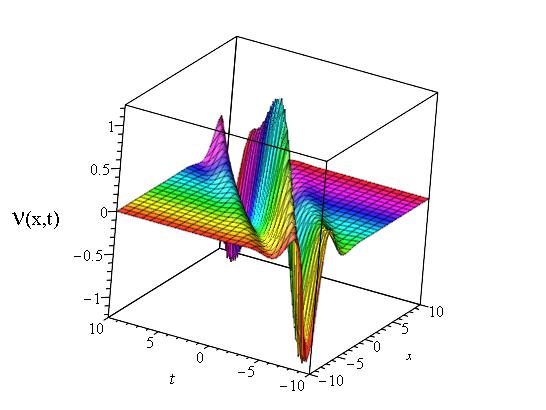}
\caption{The 2D and 3D rogue wave solitons of $\mathcal{V}_{2,2}(x,t)$, for the real part of $\mathcal{V}_{2,2}(x,t)$ with $a_1=b_1=b_2=\delta=k=\vartheta=\lambda=1$, $a_2=q=2$, $\mu=3$, $\varepsilon=0$ and $\nu=1$.} \label{fig:figure6}
\end{center}
\end{figure}

\begin{figure}[b]
\begin{center}
\includegraphics[width=6cm, height=6cm]{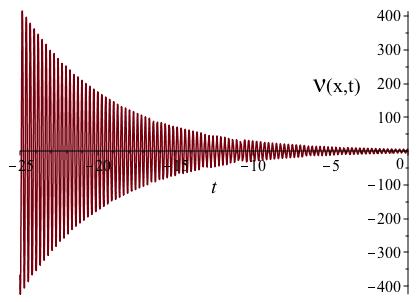}\qquad\qquad \includegraphics[width=6cm, height=6cm]{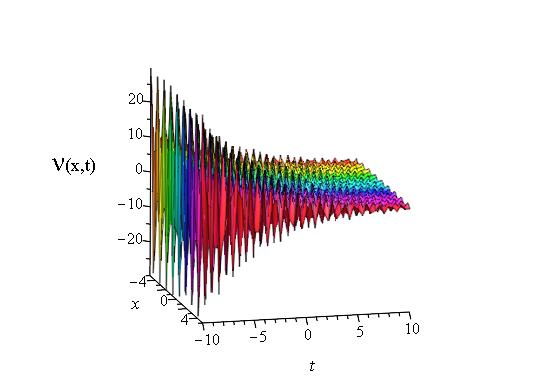}
\caption{The 2D and 3D periodic double solitons of $\mathcal{V}_{2,6}(x,t)$, for the real part of $\mathcal{V}_{2,6}(x,t)$ with $a_1=0.5$, $a_2=b_1=b_2=q=1$, $\delta=1+\sqrt{-1}$, $k=25$, $\varepsilon=0$, $\vartheta=\mu=3$ and $\lambda=\nu=0$.} \label{fig:figure7}
\end{center}
\end{figure}

\begin{figure}[b]
\begin{center}
\includegraphics[width=6cm, height=6cm]{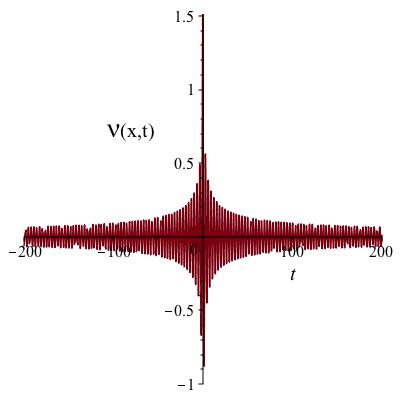}\qquad\qquad \includegraphics[width=6cm, height=6cm]{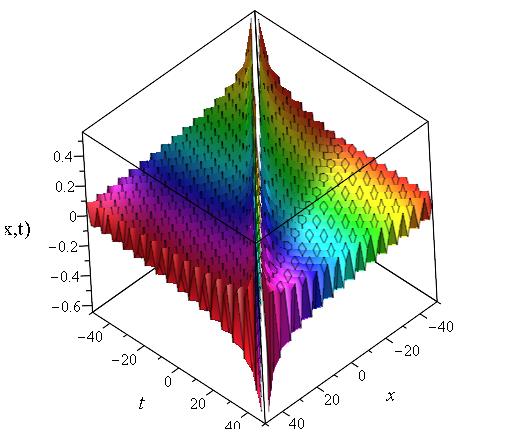}
\caption{The 2D ($x=0$) and 3D periodic double solitons of $\mathcal{V}_{2,8}(x,t)$, for the imaginary part of $\mathcal{V}_{2,8}(x,t)$ with $a_1=a_2=b_1=b_2=\delta=k=\vartheta=\mu=1$,  $\varepsilon=\lambda=0$, $q=2$ and $\nu=-1$.} \label{fig:figure8}
\end{center}
\end{figure}

The balance number of (\ref{ODESS2}) comes from power balance of the highest derivative with the highest-order nonlinear terms \cite{Bis1} that leads to $n=2$, and hence the trial solution (\ref{E3}) of the $\tan(\circleddash/2)$ expansion method takes the form,
 \begin{eqnarray}
\Psi(\circleddash)=\mathcal{L}_0 +\mathcal{L}_1 \tan(\frac{\circleddash}{2})+\mathcal{L}_2 \tan^2(\frac{\circleddash}{2}).\label{ODSS1}
\end{eqnarray}
 Putting (\ref{ODSS1}) into (\ref{ODESS2}) along with (\ref{E4}), we obtain a polynomial of $\sinh(\zeta)$, and $\cosh(\zeta)$ functions, whose coefficients after equating lead to a system of equations, and these equations enable to give us the following set of constraints:\\
 $  \delta = (4k^3q^2b_1+2\delta k\lambda^2 b_2+2\delta k\mu^2b_2-2\delta k\nu^2b_2+4k^2q^2a_1+\delta\lambda^2a_2+\delta\mu^2a_2-\delta\nu^2 a_2 -3k\lambda^2b_1-3k\mu^2b_1+3k\nu^2b_1-\lambda^2 a_1-\mu^2a_1+\nu^2a_1)/(4k^2q^2b_2+4kq^2a_2-\lambda^2b_2-\mu^2b_2+\nu^2b_2-4q^2)$,
 \newline
 \newline
  $\mathcal{L}_0 =(2\delta k^3\mu^2qb_2^2-2\delta k^3\nu^2qb_2^2+2\delta k^3\mu^2b_2^2-2\delta k^3\nu^2b_2^2+3\delta k^2\mu^2qa_2b_2-3\delta k^2\nu^2qa_2b_2-2k^3\mu^2qb_1b_2+2k^3\nu^2qb_1b_2+3\delta k^2\mu^2a_2b_2-3\delta k^2\nu^2a_2b_2+\delta k\mu^2qa_2^2-\delta k\nu^2qa_2^2-2k^3\mu^2b_1b_2 +2k^3\nu^2b_1b_2 -3k^2\mu^2qa_2b_1+3k^2\nu^2qa_2b_1-2\delta k\mu^2qb_2+\delta k\mu^2a_2^2 +2\delta k\nu^2 qb_2-\delta k\nu^2a_2^2-3k^2\mu^2a_2b_1+3k^2\nu^2a_2b_1-k\mu^2qa_1a_2 +k\nu^2qa_1a_2-2\delta k\mu^2b_2+2\delta k\nu^2b_2-\delta\mu^2qa_2+\delta\nu^2qa_2+3k\mu^2qb_1-k\mu^2a_1a_2-3k \nu^2qb_1+k\nu^2 a_1a_2- \delta\mu^2 a_2+\delta\nu^2a_2+3k\mu^2b_1-3k\nu^2b_1+\mu^2qa_1-\nu^2q a_1+\mu^2a_1-\nu^2a_1)/((\varepsilon+\vartheta)(4k^2q^2b_2+4kq^2a_2-\lambda^2b_2-\mu^2b_2+\nu^2b_2-4q^2)k)$,
\newline
 \newline
 $\mathcal{L}_1= (2\lambda(2\delta k^3\mu qb_2^2-2\delta k^3\nu qb_2^2+2\delta k^3\mu b_2^2-2\delta k^3\nu b_2^2+3\delta k^2\mu qa_2b_2 -3\delta k^2\nu qa_2b_2-2k^3\mu qb_1b_2+2k^3\nu qb_1b_2+3\delta k^2\mu a_2b_2-3\delta k^2\nu a_2b_2+\delta k\mu qa_2^2-\delta k\nu qa_2^2-2k^3\mu b_1b_2+2k^3\nu b_1 b_2-3k^2\mu qa_2b_1+3k^2\nu qa_2b_1-2\delta k\mu q b_2+\delta k\mu a_2^2+2\delta k\nu q b_2-\delta k\nu a_2^2-3k^2\mu a_2b_1+3k^2\nu a_2b_1-k\mu qa_1a_2+k\nu qa_1a_2-2\delta k\mu b_2+2\delta k\nu b_2-\delta\mu q a_2+\delta\nu q a_2+3k\mu qb_1-k\mu a_1a_2-3k\nu q b_1+k\nu a_1a_2-\delta\mu a_2+\delta\nu a_2+3k\mu b_1-3k\nu b_1+\mu q a_1-\nu q a_1+\mu a_1-\nu a_1))/(k(4k^2q^2\varepsilon b_2 +4k^2q^2\vartheta b_2+4kq^2\varepsilon a_2+4kq^2\vartheta a_2-\lambda^2\varepsilon b_2-\lambda^2\vartheta b_2-\mu^2\varepsilon b_2-\mu^2\vartheta b_2+\nu^2\varepsilon b_2+\nu^2\vartheta b-2-4q^2\varepsilon-4q^2\vartheta))$, 
 \newline
 \newline
  $ \mathcal{L}_2 = -(2\delta k^3\mu^2qb_2^2-4\delta k^3\mu\nu qb_2^2+2\delta k^3\nu^2qb_2^2+2\delta k^3\mu^2b_2^2-4\delta k^3\mu\nu b_2^2+2\delta k^3\nu^2 b_2^2+3\delta k^2\mu^2qa_2b_2-6\delta k^2\mu\nu qa_2b_2+3\delta k^2\nu^2 qa_2b_2-2k^3\mu^2qb_1b_2+4k^3\mu\nu qb_1b_2-2k^3\nu^2qb_1b_2+3\delta k^2\mu^2a_2b_2-6\delta k^2\mu\nu a_2b_2+3\delta k^2\nu^2a_2b_2+\delta k\mu^2qa_2^2-2\delta k\mu\nu qa_2^2+\delta k\nu^2qa_2^2-2k^3\mu^2b_1b_2+4k^3\mu\nu b_1b_2-2k^3\nu^2b_1b_2-3k^2\mu^2qa_2b_1+6k^2\mu\nu qa_2b_1-3k^2\nu^2qa_2b_1-2\delta k\mu^2q_2+\delta k\mu^2a_2^2 +4\delta k\mu\nu qb_2-2\delta k\mu\nu a_2^2-2\delta k\nu^2qb_2+\delta k\nu^2a_2^2-3k^2\mu^2a_2b_1+6k^2\mu\nu a_2b_1-3k^2\nu^2a_2b_1-k\mu^2qa_1a_2+2k\mu\nu qa_1a_2-k\nu^2qa_1a-2-2\delta k\mu^2b_2+4\delta k\mu\nu b_2-2\delta k\nu^2b_2-\delta\mu^2 qa_2+2\delta\mu\nu qa_2-\delta\nu^2 qa_2+3k\mu^2qb_1-k\mu^2 a_1a_2-6k\mu\nu qb_1+2k\mu\nu a_1a_2+3k\nu^2qb_1-k\nu^2a_1a_2-\delta\mu^2a_2+2\delta\mu\nu a_2-\delta\nu^2a_2+3k\mu^2b_1-6k\mu\nu b_1+3k\nu^2b_1+\mu^2qa_1-2\mu\nu qa_1+\nu^2qa_1+\mu^2a_1-2\mu\nu a_1+\nu^2a_1)/(k(4k^2q^2\varepsilon b_2+4k^2q^2\vartheta b_2+4kq^2\varepsilon a_2+4kq^2\vartheta a_2-\lambda^2\varepsilon b_2-\lambda^2\vartheta b_2-\mu^2\varepsilon b_2-\mu^2\vartheta b_2+\nu^2\varepsilon b_2+\nu^2\vartheta b_2-4q^2\varepsilon-4q^2\vartheta))$.
\\
Now using the above constraints into (\ref{ODSS1}), and combining this with $\mathcal{U}=\mathcal{T}^{\frac{1}{2q}}$ to substitute into (\ref{TRV1}), then the solutions Set [1-19] of (\ref{E4}), provides us the following exact soliton solutions of (\ref{MODS2}): 
\\
 $\mathcal{V}_{2,1}(x,t)=\left\{\mathcal{L}_0+\mathcal{L}_1\left[ \frac{\lambda}{\mu -\nu}-\frac{\sqrt{-\Lambda}}{\mu -\nu}\tan(\frac{\sqrt{-\Lambda}}{2}\bar{\zeta}) \right]\right.$ \\$\left.\qquad\qquad\quad  +\mathcal{L}_2\left[ \frac{\lambda}{\mu -\nu}-\frac{\sqrt{-\Lambda}}{\mu -\nu}\tan(\frac{\sqrt{-\Lambda}}{2}\bar{\zeta}) \right]^2\right\}^{\frac{1}{2q}} e^{i(-kx+wt+\varepsilon)} $,
\\
$\mathcal{V}_{2,2}(x,t)=\left\{\mathcal{L}_0+\mathcal{L}_1\left[ \frac{\lambda}{\mu -\nu}+\frac{\sqrt{\Lambda}}{\mu -\nu}\tanh(\frac{\sqrt{\Lambda}}{2}\bar{\zeta}) \right]\right.$ \\$\left.\qquad\qquad\quad  +\mathcal{L}_2\left[ \frac{\lambda}{\mu -\nu}+\frac{\sqrt{\Lambda}}{\mu -\nu}\tanh(\frac{\sqrt{\Lambda}}{2}\bar{\zeta}) \right]^2\right\}^{\frac{1}{2q}}e^{i(-kx+wt+\varepsilon)} $,
\\
$\mathcal{V}_{2,3}(x,t)=\left\{\mathcal{L}_0+\mathcal{L}_1\left[ \frac{\lambda}{\mu}+\frac{\sqrt{\lambda^2+\mu^2}}{\mu}\tanh(\frac{\sqrt{\lambda^2+\mu^2}}{2}\bar{\zeta}) \right]\right.$ \\$\left.\qquad\qquad\quad  + \mathcal{L}_2\left[ \frac{\lambda}{\mu}+\frac{\sqrt{\lambda^2+\mu^2}}{\mu}\tanh(\frac{\sqrt{\lambda^2+\mu^2}}{2}\bar{\zeta}) \right]^2\right\}^{\frac{1}{2q}} e^{i(-kx+wt+\varepsilon)} $,
\\
$\mathcal{V}_{2,4}(x,t)=\left\{\mathcal{L}_0+\mathcal{L}_1\left[ -\frac{\lambda}{\nu}+\frac{\sqrt{\nu^2-\lambda^2}}{\nu}\tan(\frac{\sqrt{\nu^2-\lambda^2}}{2}\bar{\zeta}) \right]\right.$ \\$\left.\qquad\qquad\quad +\mathcal{L}_2\left[ -\frac{\lambda}{\nu}+\frac{\sqrt{\nu^2-\lambda^2}}{\nu}\tan(\frac{\sqrt{\nu^2-\lambda^2}}{2}\bar{\zeta}) \right]^2\right\}^{\frac{1}{2q}}e^{i(-kx+wt+\varepsilon)}$,
\\
$\mathcal{V}_{2,5}(x,t)=\left\{\mathcal{L}_0+\mathcal{L}_1 \left[ \sqrt{\frac{\mu+\nu}{\mu-\nu}}\tanh(\frac{\sqrt{\mu^2-\nu^2}}{2}\bar{\zeta}) \right]\right.$ \\$\left.\qquad\qquad\quad  +\mathcal{L}_2 \left[ \sqrt{\frac{\mu+\nu}{\mu-\nu}}\tanh(\frac{\sqrt{\mu^2-\nu^2}}{2}\bar{\zeta}) \right]^2\right\}^{\frac{1}{2q}} e^{i(-kx+wt+\varepsilon)}$,
\\
$\mathcal{V}_{2,6}(x,t)=\left\{\mathcal{L}_0+\mathcal{L}_1\tan \left( \frac{1}{2}\tan^{-1}\left[ \frac{e^{2\mu\bar{\zeta}}-1}{e^{2\mu\bar{\zeta}}+1}, \frac{2 e^{\mu\bar{\zeta}}}{e^{2\mu\bar{\zeta}}+1} \right] \right) \right.$ \\$\left.\qquad\qquad\quad +\mathcal{L}_2\tan^2 \left( \frac{1}{2}\tan^{-1}\left[ \frac{e^{2\mu\bar{\zeta}}-1}{e^{2\mu\bar{\zeta}}+1}, \frac{2 e^{\mu\bar{\zeta}}}{e^{2\mu\bar{\zeta}}+1} \right] \right)\right\}^{\frac{1}{2q}}e^{i(-kx+wt+\varepsilon)}$,
\\
$\mathcal{V}_{2,7}(x,t)=\left\{\mathcal{L}_0+\mathcal{L}_1\tan \left( \frac{1}{2}\tan^{-1}\left[ \frac{e^{2\mu\bar{\zeta}}}{e^{2\mu\bar{\zeta}}+1}, \frac{2 e^{\mu\bar{\zeta}}-1}{e^{2\mu\bar{\zeta}}+1} \right]\right) \right.$ \\$\left.\qquad\qquad\quad   +\mathcal{L}_2\tan^2 \left( \frac{1}{2}\tan^{-1}\left[ \frac{e^{2\mu\bar{\zeta}}}{e^{2\mu\bar{\zeta}}+1}, \frac{2 e^{\mu\bar{\zeta}}-1}{e^{2\mu\bar{\zeta}}+1} \right]\right)\right\}^{\frac{1}{2q}}e^{i(-kx+wt+\varepsilon)}$,
\\
$\mathcal{V}_{2,8}(x,t)=\left\{\mathcal{L}_0+\mathcal{L}_1\left[ \frac{\lambda \bar{\zeta}+2}{(\mu-\nu)\bar{\zeta}} \right]+\mathcal{L}_2\left[ \frac{\lambda \bar{\zeta}+2}{(\mu-\nu)\bar{\zeta}} \right]^2\right\}^{\frac{1}{2q}}e^{i(-kx+wt+\varepsilon)}$,
\\
$\mathcal{V}_{2,10}(x,t)=\left\{\mathcal{L}_0+\mathcal{L}_1\tan \left( -\tan^{-1}\left[\frac{e^{k\lambda \bar{\zeta}}}{e^{k\lambda \bar{\zeta}}-1}  \right]\right) \right.$ \\$\left.\qquad\qquad\quad  +\mathcal{L}_2\tan^2 \left( -\tan^{-1}\left[\frac{e^{k\lambda \bar{\zeta}}}{e^{k\lambda \bar{\zeta}}-1}  \right]\right)\right\}^{\frac{1}{2q}}e^{i(-kx+wt+\varepsilon)}$,
\\
$\mathcal{V}_{2,11}(x,t)=\left\{\mathcal{L}_0+\mathcal{L}_1\tan \left( -\tan^{-1}\left[\frac{(\lambda+\mu)e^{\mu \bar{\zeta}}-1}{(\lambda-\mu)e^{\mu \bar{\zeta}}-1}  \right]\right)\right.$ \\$\left.\qquad\qquad\quad  +\mathcal{L}_2\tan^2 \left( -\tan^{-1}\left[\frac{(\lambda+\mu)e^{\mu \bar{\zeta}}-1}{(\lambda-\mu)e^{\mu \bar{\zeta}}-1}  \right]\right)\right\}^{\frac{1}{2q}}e^{i(-kx+wt+\varepsilon)}$,
\\
$\mathcal{V}_{2,12}(x,t)=\left\{\mathcal{L}_0+\mathcal{L}_1\left[ \frac{(\mu+\nu)e^{\mu \bar{\zeta}}+1}{(\mu -\nu)e^{\mu \bar{\zeta}}-1} \right] +\mathcal{L}_2\left[ \frac{(\mu+\nu)e^{\mu \bar{\zeta}}+1}{(\mu -\nu)e^{\mu \bar{\zeta}}-1} \right]^2\right\}^{\frac{1}{2q}}e^{i(-kx+wt+\varepsilon)}$,
\\
$\mathcal{V}_{2,13}(x,t)=\left\{\mathcal{L}_0+\mathcal{L}_1\left[ \frac{e^{\mu \bar{\zeta}}+\mu-\lambda}{e^{\mu \bar{\zeta}}-\mu-\lambda}\right] +\mathcal{L}_2\left[ \frac{e^{\mu \bar{\zeta}}+\mu-\lambda}{e^{\mu \bar{\zeta}}-\mu-\lambda}\right]^2\right\}^{\frac{1}{2q}}e^{i(-kx+wt+\varepsilon)}$,
\\
$\mathcal{V}_{2,14}(x,t)=\left\{\mathcal{L}_0+\mathcal{L}_1\left[ \frac{\lambda e^{\lambda \bar{\zeta}}}{1- \nu e^{\lambda \bar{\zeta}}}\right] +\mathcal{L}_2\left[ \frac{\lambda e^{\lambda \bar{\zeta}}}{1- \nu e^{\lambda \bar{\zeta}}}\right]^2\right\}^{\frac{1}{2q}}e^{i(-kx+wt+\varepsilon)}$,
\\
$\mathcal{V}_{2,15}(x,t)=\left\{\mathcal{L}_0+\mathcal{L}_1\tan \left( -\tan^{-1}\left[\frac{\nu \bar{\zeta}+2}{\nu \bar{\zeta}} \right]\right) +\mathcal{L}_2\tan^2 \left( -\tan^{-1}\left[\frac{\nu \bar{\zeta}+2}{\nu \bar{\zeta}} \right]\right)\right\}^{\frac{1}{2q}}e^{i(-kx+wt+\varepsilon)}$,
\\
$\mathcal{V}_{2,17}(x,t)=\left\{\mathcal{L}_0+\mathcal{L}_1\tan \left( -\tan^{-1}\left[\frac{1}{\nu \bar{\zeta}} \right]\right) +\mathcal{L}_2\tan^2 \left( -\tan^{-1}\left[\frac{1}{\nu \bar{\zeta}} \right]\right)\right\}^{\frac{1}{2q}}e^{i(-kx+wt+\varepsilon)}$,
\\
$\mathcal{V}_{2,18}(x,t)=\left\{\mathcal{L}_0+\mathcal{L}_1\tan \left[\frac{\nu \zeta +\mathcal{K}}{2} \right] +\mathcal{L}_2\tan^2 \left[\frac{\nu \zeta +\mathcal{K}}{2} \right]\right\}^{\frac{1}{2q}}e^{i(-kx+wt+\varepsilon)}$,
\\
$\mathcal{V}_{2,19}(x,t)=\left\{\mathcal{L}_0+\mathcal{L}_1\left[ \frac{e^{\lambda \bar{\zeta}}-\nu}{\lambda} \right] +\mathcal{L}_2\left[ \frac{e^{\lambda \bar{\zeta}}-\nu}{\lambda} \right]^2\right\}^{\frac{1}{2q}}e^{i(-kx+wt+\varepsilon)}$,\\
where $\mathcal{L}_0$, $\mathcal{L}_1$ and $\mathcal{L}_2$ come from the above set of constraints. There is a note that the solutions $\mathcal{V}_{2,9}(x,t)$ and $\mathcal{V}_{2,16}(x,t)$ are invalid for the constraints.
 \newline
 \newline
To the graphical representations, it is shown that $\mathcal{V}_{2,1}(x,t)$, $\mathcal{V}_{2,4}(x,t)$, $\mathcal{V}_{2,18}(x,t)$ solutions provide double period optical solitons. In particular, we present the $2D$ and 3D graphs in Figure \ref{fig:figure5} for the solution $\mathcal{V}_{2,1}(x,t)$; and the solutions  $\mathcal{V}_{2,4}(x,t)$ and $\mathcal{V}_{2,18}(x,t)$ have similar dynamical characteristics to the Figure \ref{fig:figure5}.
\\
The solutions $\mathcal{V}_{2,2}(x,t)$, $\mathcal{V}_{2,3}(x,t)$ and $\mathcal{V}_{2,5}(x,t)$ represent interaction between periodic waves and solitons which produce an optical double solitons. We present graphs in Figure \ref{fig:figure6} for a particular solution $\mathcal{V}_{2,3}(x,t)$. It is shown that the solution presents rogue waves whose amplitude increased two or three times higher than the surrounding waves within a tiny time. These types of rogue waves can be predicted to control the amplitude of the waves.
\\
The solutions $\mathcal{V}_{2,6}(x,t)$, $\mathcal{V}_{2,7}(x,t)$ $\mathcal{V}_{2,10}(x,t)$, $\mathcal{V}_{2,11}(x,t)$, $\mathcal{V}_{2,12}(x,t)$, $\mathcal{V}_{2,13}(x,t)$, $\mathcal{V}_{2,14}(x,t)$ and $\mathcal{V}_{2,19}(x,t)$ provide to similar interaction of periodic wave with optical solitons. We give graphs in Figure \ref{fig:figure7} for the solution $\mathcal{V}_{2,6}(x,t)$. The graphs show that the amplitude of the optical solitons gradually decreases. 
\\
The solutions $\mathcal{V}_{2,8}(x,t)$,  $\mathcal{V}_{2,15}(x,t)$, and $\mathcal{V}_{2,17}(x,t)$  provide to similar interaction of periodic wave with optical solitons. We present a periodic  double solitons for the solution $\mathcal{V}_{2,8}(x,t)$ in Figure \ref{fig:figure8}. The graphs show that the amplitude of the wave gradually increases and reaches the highest peak at the moment of interaction, and then the amplitude goes to diminish for a larger time.  
\\
{\bf Remark.} When $\mathcal{L}_{2}=0$ and $q=1/2$ in the obtained solutions, then the corresponding results in sections 4.2 and 3.2 are making identical both in mathematically and dynamically. For the large values of $q$, it is shown that the amplitude of the waves increases more, and the waves can be transferred to a very large distance remaining same nonlinear dynamical shapes.

\section{Conclusion} 
The main results of this paper is the determination of exact solitons to the Biswas-Arshed model (BAM) with nonlinear Kerr and power law via the $\tan(\circleddash/2)$ expansion integral scheme. We retrieve optical shock waves, double periodic optical solitons, interaction between optical periodic waves and optical solitons, and optical periodic waves and rogue waves of the model. We explain how to display the nonlinear feathers of the waves at the moment of interactions, that is, how to the amplitude of the optical periodic double solitons gradually increases, reaches to the highest peak, and goes to disappear for a much longer time. In fact, all these types of optical solitons can be frequently used to amplify or reduce waves on account of a certain hight. The physical phenomena of the solitons are graphically presented in the 2D and 3D plots. To our knowledge  these types of solitons for the Biswas-Arshed model have not been explored before \cite{Yil1, Tah1, Tah2, Reh1}. In fact, the model could be investigated to construct multi- and rogue waves solitons by the other existing methods, in particular, Hirota bilinear approach \cite{Ros1, Ros2} and Darboux transformation \cite{Bag1}.

\end{document}